\documentclass[aps,twocolumn,superscriptaddress,altaffilletter,lengthcheck,
tightenlines,showpacs,showkeys]{revtex4}

\newcommand{\al}{\alpha}
\newcommand{\bb}{\beta}
\newcommand{\D}{\Delta}
\newcommand{\ben}{\begin{eqnarray}}
\newcommand{\een}{\end{eqnarray}}
\newcommand{\be}{\begin{equation}}
\newcommand{\ee}{\end{equation}}
\newcommand{\ba}{\begin{eqnarray}}
\newcommand{\ea}{\end{eqnarray}}
\newcommand{\n}{\label}

\newcommand{\ga}{\gamma}
\newcommand{\ro}{\rho}

\newcommand{\bn}{\begin{equation}\label}
\newcommand{\m}{MHRDE }

\usepackage[dvipdf]{epsfig}
\usepackage{color}

\begin{document}

\title{Interacting dark matter and modified holographic Ricci dark energy \\
induce a relaxed Chaplygin gas  }

\author{Luis P. Chimento}\email{chimento@df.uba.ar}
\affiliation{Departamento de F\'{\i}sica, Facultad de Ciencias Exactas y Naturales,  Universidad de Buenos Aires, Ciudad Universitaria, Pabell\'on I, 1428 Buenos Aires, Argentina}
\author{Mart\'{\i}n G. Richarte}\email{martin@df.uba.ar}
\affiliation{Departamento de F\'{\i}sica, Facultad de Ciencias Exactas y Naturales,  Universidad de Buenos Aires, Ciudad Universitaria, Pabell\'on I, 1428 Buenos Aires, Argentina}

\date{\today}
\bibliographystyle{plain}


\begin{abstract}
We investigate a model of interacting dark matter and dark energy given by a modified holographic Ricci cutoff in a spatially flat Friedmann-Robertson-Walker (FRW) space-time. We consider a nonlinear interaction consisting of a significant rational function of the total energy density and its first derivative homogeneous of degree one and show that the effective one-fluid obeys the equation of state of a relaxed Chaplygin gas. So that, the universe is dominated by pressureless dark matter at early times and undergoes an accelerated expansion in the far future driven by a strong negative pressure. We apply the $\chi^{2}$-statistical method to the observational Hubble data and the Union2 compilation of SNe Ia for constraining the cosmological parameters and analyze the feasibility of the modified holographic Ricci cutoff. By using the new  $Om$ diagnostic method, we find that the effective model differs substantially from the $\Lambda$--CDM one, because it gets the accelerated expansion faster than the $\Lambda$--CDM model. Finally, a new model with a third component decoupled from the interacting dark sector is presented for studying bounds on the dark energy at early times.

\end{abstract} 
\vskip 1cm


\bibliographystyle{plain}

\maketitle


\section{Introduction}
The holographic principle states that  the maximum  number of degrees of freedom  in a volume should be proportional to the surface area \cite{holo1}, \cite{holo2}, \cite{holo3}, \cite{holo4}.  Using the effective  quantum field theory  it was shown that  the zero-point  energy  of a system  with size $L$ should no exceed  the mass  of a black hole with the same size, thus $L^{3}\ro_{\Lambda}\leq LM^{2}_{P}$, where $\ro_{\Lambda}$ corresponds to the quantum zero-point energy density \cite{holo5} and $M^{-2}_{P}= 8\pi G$. The latter relation establishes a link between  the ultraviolet cutoff, define through $\ro_{\Lambda}$, and the infrared cutoff  which is encoded by the scale $L$. By applying this novel principle within the cosmological context, one  can assume that the dark energy density of the universe $\ro_{x}$ takes the same form of the vacuum energy, $\ro_{\Lambda}=\ro_{x}$. Taking the largest $L$ as the one saturating the above inequality, it turns out to be the holographic dark energy is given by $\rho_{x}=3c^{2}M^{2}_{~P}L^{-2}$, where $c$ is a numerical factor. Here, the infrared cutoff will be  considered  as a function of the cosmic time so the holographic dark energy will evolve dynamically and in order to be taken as a good candidate  for occupying  the place of  the cosmic fuel called dark energy it should be able to mimic the current accelerated expansion phase of the Universe. So far in the literature, the IR cutoff   has been  taken  as the large scale of the universe, Hubble horizon \cite{hde0,hde1}, particle horizon, event horizon \cite{hde1} or generalized IR cutoff \cite{ricci1}, \cite{ricci2},  \cite{go}, \cite{odi1}, \cite{odi2}. 

An interesting  property of the modified Ricci cutoff \cite{go}  refers to the fine-tuning problem, that is, due to the holographic dark energy density is based on spacetime scalar curvature without involving a Planck or high physical energy scale the fine-tuning problem might be avoided, further the coincidence problem seems to disapear within this context \cite{ricci1}. For  all these  reasons, we will focus  on a holographic model where the dark energy density corresponds to the modified Ricci  cutoff \cite{go}, \cite{go2a}, \cite{go2b} \cite{HL1}, \cite{HL2}  extending  the recently investigated linear interaction model  \cite{HL2} to the case of  nonlinear ones.

Another central aspect  of the modern cosmology  is associated  with the exchange of energy between the dark components, i.e., the dark matter not only can feel the presence of  the dark energy through a  gravitational expansion of the Universe but also  can interact between them.  Because the nature of the dark sector still remains  unknown there is no a fundamental reason to specify the coupling between dark energy and dark matter. In fact, one follows a phenomenological approach by  studying the properties hidden in each type of  interaction  and then one confronts  the theoretical model with the available observational data. Reference \cite{jefe1} investigated several linear and nonlinear interactions in the dark sector, generalizing previous works on the literature. It was introduced  an effective one-fluid description  of the dark components and shown  that interacting  and unified models  are related to each other. Also, a generic  nonlinear interaction induces an effective equation of state  resembling a variable Chaplygin gas model, giving rise to the ``relaxed Chaplygin gas model'' \cite{jefe1}.  Within this framework, we are going  to  explore an holographic model and its relation to relaxed  Chaplygin gas.

The paper is outlined as follows. In Sec. II we introduce the model and discuss about the  exchange of energy in the dark sector where the modified Ricci cutoff is taken into account. For the selected nonlinear interaction term, we show that the effective one-fluid obeys the equation of state of a relaxed Chaplygin gas. In Sec. III we find the most probable value of the cosmological parameters by applying the $\chi^{2}$ minimization method to the  Hubble data and  the Union2 compilation of SNe Ia. We  use the best--fit value of the parameters for confronting our model with the standard $\Lambda$--CDM and apply the $Om$ diagnostic method. Finally, in Sec.IV we summarize our findings. 

\section{The model }

We present an holographic cosmological model, where we associate the IR cutoff with the modified Ricci's radius and take $L^{-2}$ in the form of a linear combination of $\dot H$ and $H^2$. After that, the modified holographic Ricci dark energy (MHRDE) \cite{go}, $\rho_{x}=3c^{2}M^{2}_{~P}L^{-2}$, becomes 
\be
\n{03}
\ro_x=\frac{2}{\al -\bb}\left(\dot H + \frac{3\al}{2} H^2\right).
\ee
Here $H = \dot a/a$ is the Hubble expansion rate, $a$ is the scale factor and $\al, \bb$ are free constants. Introducing the variable $\eta = \ln(a/a_0)^{3}$, with $a_0$ the present value of the scale factor and $' \equiv d/d\eta$, the above MHRDE (\ref{03}) becomes a modified conservation equation (MCE) for the cold dark matter $\ro_c$ and the MHRDE
\be
\n{06}
\ro'=-\al\ro_c -\bb\ro_x,
\ee
after using the Friedmann equation, 
\be
\n{00}
3H^2= \ro_c + \ro_x,
\ee 
for a spatially flat FRW cosmology and $\ro=\ro_c + \ro_x$. The MCE (\ref{06}) looks as it were a conservation equation for both dark components with constant equations of state. In connection with observations on the large scale structures, which seems to indicate that the Universe  must have been dominated by nearly pressureless components, we assume that $\ro_c$ includes all these components and has an equation of state $p_c=0$ while the \m has a barotropic index $\omega_x=p_x/ \ro_x$, so that the whole conservation equation (WCE) becomes 
\be
\n{05}
\ro'=-\ro_c -(\omega_x +1)\ro_x.
\ee
The compatibility between the MCE (\ref{06}) and the WCE (\ref{05}) yields a linear dependence of the equation of state of the MHRDE 
\be
\n{07}
\omega_x=(\al -1)r+ (\beta-1),
\ee
with the ratio of both dark components $r=\ro_c/\ro_x$. Solving the linear algebraic system of equations (\ref{06}) and $\ro=\ro_c+\ro_x$ we obtain both dark energy densities as functions of $\ro$ and $\ro'$
\be
\n{10}
\ro_c= - \frac{\bb \ro +\ro '}{\D\ga}, \qquad \ro_x= \frac{\al \ro +\ro '}{\D\ga},
\ee
with $\D\ga = \al -\bb$, while the total pressure $p = p_c + p_x$ is given by 
\be
\n{12}
p(\ro,\ro') = -\ro -\ro',
\ee
 
From now on we will use the MCE (\ref{05}) instead of the WCE with variable $\omega_{x}$ because it is simpler, and introduce an interaction between both dark components through the term $3HQ_M$ into the MCE (\ref{05}) with constant coefficients, so 
\be
\n{08}
\ro_c' + \al \ro_c = - Q_M,
\ee
\be
\n{09}
\ro_x' + \bb \ro_x = Q_M.
\ee
Finally, from Eqs. (\ref{10}) and (\ref{08}), we obtain the source equation \cite{jefe1} for the energy density $\ro$
\be
\n{14}
\ro''+(\al + \bb)\ro' + \al\bb\ro =  Q_M\D\ga.
\ee
 
Now we consider cosmological models where the interaction $Q_M$ between both dark components is nonlinear and includes a set of terms  which are homogeneous of degree 1 in the total energy density and its first derivative \cite{jefe1}, 
\be
\n{Q}
Q_M=\frac{(\al\beta -1)}{\Delta\gamma}\,\ro+\frac{(\al + \beta -\nu-2)}{\Delta\gamma}\,\ro'-\frac{\nu\ro'^{2}}{\ro\Delta\gamma},
\ee
where $\nu$ is a positive constant that parameterizes the interaction term $Q_M$. Replacing (\ref{Q}) into (\ref{14}) it turns into a nonlinear second order differential equation for the energy density 
\be
\n{Ia}
\ro\ro''+(2+\nu)\ro\ro'+\nu\ro'^{2}+\ro^2=0.
\ee
Introducing the new variable $y=\ro^{(1+\nu)}$ into the latter equation one gets a second order linear differential equation $y''+(2+\nu)y'+(1+\nu)y=0$, whose solutions allow us to write the energy density
\be
\n{23}
\ro=\left[\ro_{10}a^{-3}+\ro_{20}a^{-3 (1+\nu)}\right]^{1/(1+\nu)},  
\ee
where $\ro_{10}$ and $\ro_{20}$ are positive constants. From Eqs. (\ref{10}) and (\ref{12}) we have both dark energy densities and the total pressure
\be
\n{cI}
\ro_c=\frac{-\ro}{\al-\bb}\left[\bb-1+\frac{\nu}{(1+\nu)(1+\ro_{20}a^{-3\nu}/\ro_{10})}\right],\,\,\,
\ee
\be
\n{xI}
\ro_x=\frac{\ro}{\al-\bb}\left[\al-1+\frac{\nu}{(1+\nu)(1+\ro_{20}a^{-3\nu}/\ro_{10})}\right],\,\,\,\,\,\,\,\,\,\,\,\,	
\ee
\be
\n{24}
p=-\frac{\nu\ro_{10}}{1+\nu}\,\frac{a^{-3}}{\ro^\nu}.
\ee
From these equations we see that an initial model of interacting dark matter and dark energy can be associated with 
an effective one-fluid description of an unified cosmological scenario where the effective one-fluid, with energy density $\ro=\ro_c+\ro_x$ and pressure (\ref{24}), obeys the equation of state of a relaxed Chaplygin gas $p=b\ro+f(a)/\ro^\nu$, where $b$ is a constant \cite{jefe1}. The effective barotropic index $\omega=p/\ro=\omega_x\ro_x/\ro$ reads,
\be
\n{eb}
\omega=-\frac{\nu\ro_{10}}{(1+\nu)(\ro_{10}+\ro_{20}a^{-3\nu})}.
\ee

At early times and for $\nu>0$,  the effective energy density (\ref{23}) behaves as $\ro\approx a^{-3}$, the effective barotropic index (\ref{eb}) $\ga\approx 1$ and  the effective fluid describes an Universe dominated by nearly pressureless dark matter. However, a late time accelerated Universe i.e., $\omega<-1/3$ with positive dark energy densities require that $\nu>1/2$, $\beta<1$ and $\al>1$. From now on we adopt the latter restrictions.
 
In the case $\nu=1$, one gets an equation of state (\ref{24}) similar to a variable Chaplygin gas as the one studied in \cite{zy1}, \cite{zy2}. In these works the authors have used  the gold sample of type Ia supernova data  as well as the x-ray  gas mass fractions in galaxy clusters to obtain some observational constraints of the model.


\section{Observational Hubble data  constraints}
Now, we are going to  find the observational constraints on the parameter space of the MHRDE using the observational  $H(z)$ data. The function $H(z)$ plays a crucial role in understanding the properties of the dark energy, since its value is directly obtained from the cosmological observations. Using the absolute ages of passively evolving galaxies observed at different redshifts, one obtains the differential ages $dz/dt$ and the function $H(z)$ can be measured through the relation $H(z)=-(1+z)^{-1}dz/dt$. We take the data coming from the Gemini Deep Deep Survey \cite{obs1}, archival data \cite{obs2a}, \cite{obs2b}, \cite{obs2c}, \cite{obs2d}, \cite{obs2e}, \cite{obs2f}, Simon et al \cite{obs3} and compare it with the $H(z)$ to attempt to obtain the constraints for the current cosmological models. The 12 observational  $H(z)$ data from \cite{obs4} are listed in Table I. There, $H_{obs}(z_i)$ and $H_{obs}(z_k)$ are uncorrelated because they are obtained from the observations of galaxies at different redshifts.

The probability distribution for the $\theta$--parameters is given by a Gaussian density (see e.g.\cite{Press})
$$P(\theta)=\mathcal Z e^{-\chi^2 (\theta)/2},$$
\be
\n{c1}
\chi^2(\theta) =\sum_{k=1}^{12}\frac{[H(\theta,z_k) - H_{obs}(z_k)]^2}{\sigma(z_k)^2},
\ee
where $\mathcal Z$ is a normalization constant, $H_{obs}(z_k)$ is the observational $H(z)$ data at the redshift $z_k$, $\sigma(z_k)$ is the corresponding $1\sigma$ uncertainty, and the summation is over the 12 observational  $H(z)$ data of Table I. Since we are interested in obtaining  the bounds for the model parameters, we have adopted as a prior  $H_{0}=72.2 \pm 3.6~{\rm km~s^{-1}\,Mpc^{-1}}$ \cite{H0}. The  Hubble expansion of the model  becomes: 
\be
\n{Ht}
H(\theta; z)=H_0\Big\{B(1+z)^{3}+ (1-B)(1+z)^{3(\nu+1)}\Big\}^{\frac{1}{2(\nu+1)}}
\ee
\be
\n{B}
B[\theta]=\frac{\nu+1}{\nu}\left[\al(\Omega_{x0}-1) + (1-\beta\Omega_{x0}) \right]
\ee
where $\theta=\{\al, \beta, \Omega_{x0}, \nu\}$  and we have used that $\ro_{02}/\ro_{01}=(B-1)/B$. The two independent parameters $\al$ and $\beta$ will be fixed along the statistic analysis. Then, for a given pair of $(\al_{f}, \beta_{f})$,  we are going to perform the statistic analysis by minimizing the $\chi^2$ function to  obtain the best fit values of  the random variables $\theta_{c}=\{\nu, \Omega_{x0} \}$ that correspond to a maximum of Eq.(\ref{c1}). More precisely, the  best--fit parameters $\theta_{c}$ are those values where $\chi^2_{min}(\theta_{c})$ leads to the local minimum of the $\chi^2(\theta)$ distribution. If $\chi^2_{dof}=\chi^2_{min}(\theta_{c})/(N -n) \leq 1$ the fit is good and the data are consistent with the considered model $H(z;\theta)$. Here, $N$ is the number of data and $n$ is the number of parameters \cite{Press}. The variable $\chi^2$ is a random variable that depends on $N$ and its probability distribution is a $\chi^2$ distribution for $N-n$ degrees of freedom.

\begin{figure}[h!]
\begin{center}
\includegraphics[height=6cm,width=7.5cm]{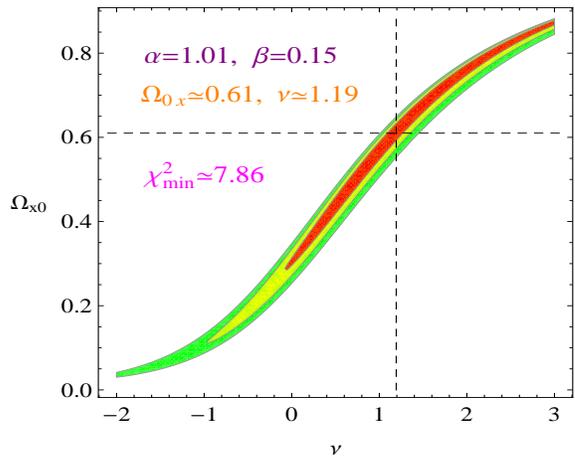}
\caption{ The confidence  contours  associated with the $1\sigma$, $2\sigma$ and $3\sigma$ error bars are shown  in the plane $(\nu, \Omega_{x0})$. The point $(\nu,\Omega_{x0})=(1.19, 0.61)$ shows the best--fit for the  $H_{obs}(z)$ data.}
\label{fig1}
\end{center}
\end{figure}

Besides, $68.3\%$ confidence  contours  in the $(\nu, \Omega_{x0} )$ plane  are made of the random data sets that satisfy the inequality $\Delta\chi^{2}=\chi^2(\theta)-\chi^{2}_{min}(\theta_{c})\leq 2.30$. The latter equation defines a bounded region by a closed area around $\theta_{c}$ in the two-dimensional parameter plane, thus the $1\sigma$ error bar can be identified with the distance from the $\theta_{c}$ point to the boundary of the  two-dimensional parameter plane. It can be shown that $95.4\%$ confidence contours  with a $2\sigma$  error bar in the samples satisfy $\Delta\chi^{2}\leq 6.17$ while the data within $99.73\%$ confidence  contours  with a $3\sigma$ error bar are accommodated in the domain defined by $\Delta\chi^{2}\leq 11.8$. After performing this analysis  we are in position to get  confidence  contours in the $(\nu, \Omega_{x0} )$ plane, thus using the  $\chi^2(\al_{f}, \beta_{f}, \nu, \Omega_{x0} $) distribution  one can find the $68.3\%$, $95.4\%$, and $99.73\%$ confidence  contours respectively [see Fig.(\ref{fig1})]. We have taken the point of reference $(\al_{f},\beta_{f})=(1.01,0.15)$ but it is possible to show a wide set of admissible values for $\al$ and $\beta$ which leads to a good fit [see Table(\ref{VP}) ]. Thus, from Fig.(\ref{fig1}) we get the best fit at $\theta_c=(\nu,\Omega_{x0})=(1.19 \pm 0.12; 0.61 \pm 0.02 )$. It corresponds to a local minimum $\chi^2_{min}=7.86$ leading to a good fit with $\chi^2_{dof}=0.786$ per degree of freedom. The Ricci's  cutoff $(\al, \beta)=(4/3,1)$ does not guarantee the convergence of the minimization process. However, the values of ($ \nu,\Omega_{x0}$) obtained from an holographic dark energy $\ro_{x} \propto R$, namely $(4/3,\beta)$, fulfills the goodness condition $\chi^2_{dof}<1$.  The values of $\Omega_{x0}$, which varies from, $0.58$ to $0.69$, do not deviate significantly from the observational limits provided by the WMAP-7 project \cite{WMAP7} with  $\Omega_{x0}=0.73$ [see Table (\ref{VP})]. Comparing the Ricci model with the one arising from MHRDE for $(\alpha=1.01,\beta=0.15)$, the former gives  $(\nu,\Omega_{x0})=( 1.19, 0.69)$, whereas the latter yields $(\nu,\Omega_{x0})=( 1.19, 0.61)$,  so the Ricci model seems to be statistically favored by $H(z)$ data  showing a $\Omega_{x0}$ closer to the observational bound reported by  the WMAP-7 project \cite{WMAP7}. 

\begin{figure}[h!]
\begin{center}
\includegraphics[height=6cm,width=7.5cm]{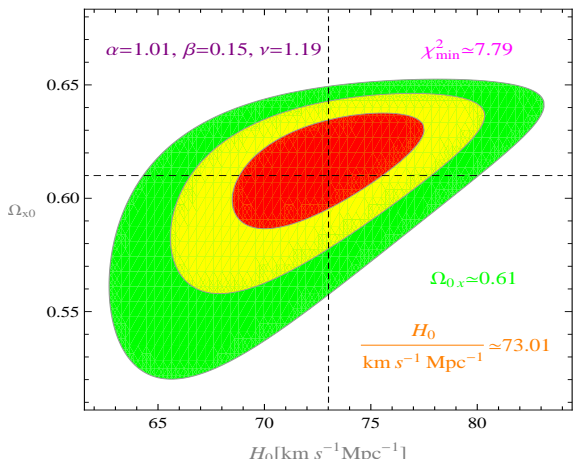}
\caption{ We show the confidence contour in the $H_{0}-\Omega_{x0}$  plane for the MHRDE model considering as priors $\nu=1.19$, $\al=1.01$ and $\beta=0.15$. The best fit value of  $(H_{0},\Omega_{x0})$ is used to compare with the $\Lambda$CDM model. }
\label{1b}
\end{center}
\end{figure}

We  estimate the best value of $H_{0}$ and $\Omega_{x0}$ for the $\Lambda$--CDM model using the Hubble data as well as the Union2 data for SNe Ia \cite{Union2}.  The former dataset  leads to $H_{0}= 73.60 \pm 3.18 ~{\rm km~s^{-1}\,Mpc^{-1}}$ and $\Omega_{x0}= 0.730 \pm 0.04$ with $\chi^2_{dof}=0.770$ whereas the latter one gives $H_{0}= 70 ~{\rm km~s^{-1}\,Mpc^{-1}}$ and $\Omega_{x0}= 0.73 $ along with $\chi^2_{dof}=0.978$. Now, in oder to make possible a comparison with the $\Lambda$--CDM model we need to estimate the same types of parameters so we take as priors $\nu=1.19$, $\al=1.01$, and $\beta=0.15$ but we allow $H_{0}$ and $\Omega_{x0}$ as free parameters to be found under the minimization process. It turns out that $H_{0}= 73.011 \pm 2.97 ~{\rm km~s^{-1}\,Mpc^{-1}}$ and  $\Omega_{x0}=0.617 \pm 0.001$ with $\chi^2_{dof}=0.779<1$ [see Fig.(\ref{1b})], then both models give cosmological bounds of the pair  $(H_{0}, \Omega_{0x})$ very consistent with the those reported in \cite{WMAP7}.

In order to compare  the Hubble data (12 points) with the Union2 compilation of 557 SNe--Ia \cite{Union2}  we proceed as follows; thus, we
took as priors $H_{0}=72.2~{\rm km~s^{-1}\,Mpc^{-1}}$, $\al=1.01$ and $\beta=0.15$  in both cosmological data. We found  the best--fit values of $\nu$ and $\Omega_{x0}$ for both sets, focusing on the existence of  some tighter constraints coming from the SNe Ia data. For the Hubble data we obtained $\al=1.19$ and $\Omega_{x0}=0.61$ with $\chi^2_{dof}=0.786$ whereas the SNe Ia data lead to $\nu=1.5$ and $\Omega_{x0}=0.70$  with $\chi^2_{dof}=0.812 <1$; in broad terms the tighter constraints seems to be found with the Hubble data. Of course, these results can vary according to the parameter regions taken into account in the minimization process.

\begin{table}[htbp!]
\begin{center}
\begin{tabular}{r|c|c}
\hline\hline
$z_i$&$H_{obs}(z_i)$&$1\sigma$\\
{redshift}&${\rm km~s^{-1}\,Mpc^{-1}}$&uncertainty \\
\hline\hline
0.00&74.2&$\pm3.6$\\
0.10&69&$\pm12$\\
0.17&83&$\pm8$\\
0.27&77&$\pm14$\\
0.40&95&$\pm17$\\
0.48&97&$\pm60$\\
0.88&90&$\pm40$\\
0.90&117&$\pm23$\\
1.30&168&$\pm17$\\
1.43&177&$\pm18$\\
1.53&140&$\pm14$\\
1.75&202&$\pm40$\\
\hline\hline
\end{tabular}
 \end{center}
    \caption{\label{Hubbledata} The observational $H(z)$ data~\cite{obs4}.}
\end{table}

\begin{table}[htbp!]
\begin{center}
\begin{tabular}{r|c|c}
\hline\hline
$(\al, \beta)$&$(\Omega_{0x}\pm \sigma, \nu \pm \sigma)$&$\chi^{2}_{dof}$\\
&$$&$$ \\
\hline\hline
(1.01,0.05)&$(0.58\pm 0.20, 1.19 \pm 1.13)$&0.786\\
(1.01,0.1)&$(0.58\pm 0.23, 1.19 \pm 1.13)$&0.786\\
(1.01, 0.15)&$(0.61 \pm 0.02, 1.19 \pm 1.13)$&0.786\\
(1.2,-0.1)&$(0.55 \pm 0.16, 1.19 \pm 1.13)$&0.786\\
(1.2,-0.05)&$(0.57 \pm 0.17, 1.19 \pm 1.13)$&0.786\\
(4/3,-0.1)&$(0.59 \pm 0.15, 1.19 \pm 1.13)$&0.786\\
(4/3,0.1)&$(0.69 \pm 0.17, 1.19 \pm 1.13)$&0.786\\
\hline\hline
\end{tabular}
 \end{center}
    \caption{\label{VP} We show the observational bounds for the pair $(\Omega_{x0}, \nu)$ varying $(\al,\beta)$ with a given prior of $H_{0}=72.2 \pm 3.6~{\rm km~s^{-1}\,Mpc^{-1}}$}
\end{table}

\begin{figure}[h!]
\begin{center}
\includegraphics[height=6cm,width=7.5cm]{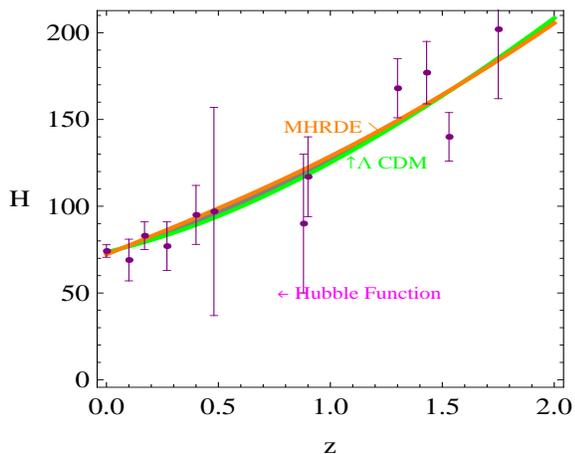}
\caption{ We plot $H(z)$ for the $\Lambda$CDM with $H_0 = 73.60$ km/s Mpc and MHRDE model with prior $H_0 = 72.22$ km/s Mpc. We also show the observational data $H(z_{k})$ with their error bars.}
\label{f3}
\end{center}
\end{figure}
\begin{figure}
\begin{center}
\includegraphics[height=6cm,width=7.5cm]{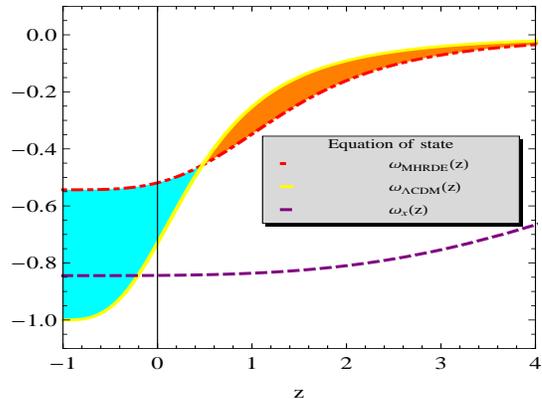}
\caption{We show the equations of state  for the effective fluid $\omega(z)$ and the dark energy $\omega_{x}(z)$.}
\label{fig5}
\end{center}
\end{figure}

Now, using the best--fit model parameters $\theta_c=(\nu,\Omega_{x0})=(1.19 \pm 1.13, 0.61 \pm 0.02 )$ we would like to compare the   model having a MHRDE with the standard $\Lambda$--CDM scheme composed of baryonic matter and a constant dark energy $\Omega_{0x}=0.73 \pm  0.04$. In Fig.(\ref{f3}) we show  
the $H_{obs}(z)$ data with their error bars, and the theoretical curves corresponding to the $\Lambda$--CDM and MHRDE models,  respectively. At small redshift $0\lesssim z\lesssim 0.2$ both models seem to be in excellent agreement with the observational $H(z)$ data,  while at large redshift $z>0.2$ the model shows a slight difference with the $\Lambda$--CDM. However, the observational data show that in the interval $z\in [0, 1]$ both models have a good adjustment. As $-1\leq\omega(z),\omega_{x}(z)\leq 0$, the equations of state of the effective fluid and dark energy do not cross the phantom line, at least for the best--fit model parameters used previously [see Fig.(\ref{fig5})]. The expression of $\omega_{x}$ at present 
\be
\n{wx0}
\omega_{x0}= -\frac{\nu(\alpha -\beta)B}{(\alpha-1)(1+\nu)+\nu B}, 
\ee
becomes $\omega_{x0}=-0.88$ when evaluating at the best--fit values $\theta_c=(\nu,\Omega_{x0}, \alpha, \beta)=(1.19, 0.61, 1.01, 0.15)$. It is  close to the value reported by WMAP-7, $\omega_{x0}=-0.93$,  when the joint analysis of the WMAP+BAO+$H_{0}$+SN data \cite{WMAP7} for constraining the present-day value of the equation of state for dark energy is made.

\begin{figure}
\begin{center}
\includegraphics[height=6cm,width=7.5cm]{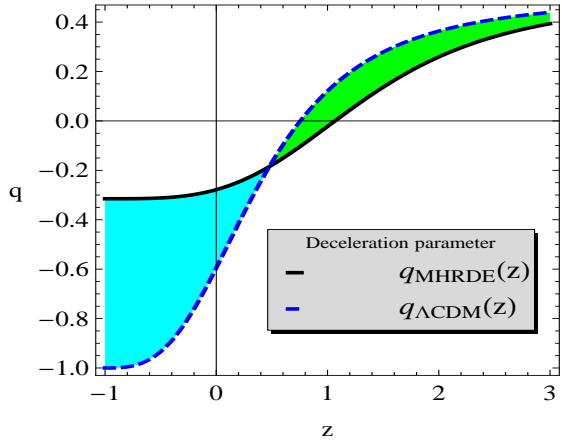}
\caption{The deceleration parameter  $q(z)$.}
\label{fig6}
\end{center}
\end{figure}

Fig. (\ref{fig6}) shows the evolution of the decelerating parameter $q=-\ddot{a}/aH^{2}$ with the redshift $z$ for the MHRDE and $\Lambda$--CDM models.  It takes the form 
\be
\n{q0}
q_{0}= \frac{1+ \nu(1-3B) }{2(1+\nu)}, 
\ee
at $z=0$ for the former model. Using the best--fit values $\theta_c=(\nu,\Omega_{x0}, \alpha, \beta)=(1.19, 0.61, 1.01, 0.15)$ in Eq. (\ref{q0}), one gets  $q_{0}=-0.27$ while for the $\Lambda$-CDM model one obtains $q_{0}=-0.59$. The critical redshift where the acceleration starts, 
\be
\n{za}
z_{acc}= -1+ \Big[\frac{(2\nu-1)B}{(1+\nu)(1-B)}\Big]^{1/3\nu}, 
\ee
turns to be $z_{acc}=1.06$ for the best--fit values $\theta_c$, then our model enters the accelerated regime earlier than the  $\Lambda$CDM one with $z_{acc}=0.75$.

\begin{figure}
\begin{center}
\includegraphics[height=6cm,width=7.5cm]{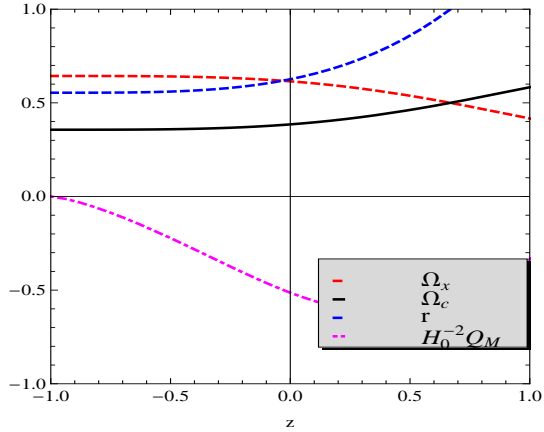}
\caption{The density parameters ($\Omega_x$, $\Omega_c$), the ratio $r=\Omega_c/\Omega_x$, and $H^{-2}_{0}Q_{M}$ are shown versus the redshift $z$.}
\label{fig7}
\end{center}
\end{figure}

In Fig.(\ref{fig7}) we plot the density parameters $\Omega_c$, $\Omega_x$, its ratio $r(z)$ and find the present-day values of $\Omega_{x0}=0.61$, $\Omega_{c0}=0.39$ and $r=0.62$. It shows that the  model with a MHRDE seems to be  appropriated for resolving the coincidence problem. Regarding the  modified Ricci coupling function,  one can show that $Q_{M}\leq 0$  and the coupling decreases its strength with the redshift and goes to zero in the far future, $z \rightarrow -1$ [see  Fig.(\ref{fig7})].

\begin{figure}
\begin{center}
\includegraphics[height=6cm,width=7.5cm]{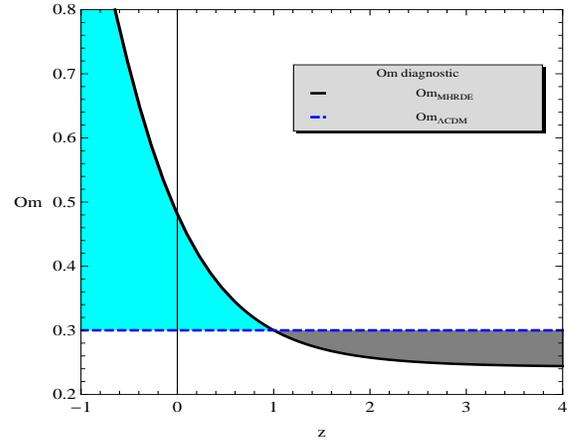}
\caption{The  $Om(z)$ diagnostic.}
\label{fig8}
\end{center}
\end{figure}

So far, we have discussed some cosmological constraints coming from the observational $H(z)$ data at different redshifts. Let us take into account an alternative  method, which is beyond parameter constraints, that can be implemented to distinguish the $\Lambda$--CDM from other dark energy models. The  new diagnostic of dark energy  called $Om(z)$ is an interesting one because it does not involve the cosmic equation of state \cite{om}. The $Om(z)$ function is constructed from the Hubble parameter $H$ and  defined as \cite{om}
\be
\n{om}
Om(z)=\frac{H^2-H^2_{0}}{H^2_{0}[(z+1)^{3}-1]},
\ee
Since the $Om(z)$ only depends upon the scale factor and its first derivative, one considers the present diagnostic method as a geometrical one. Futhermore, for the $\Lambda$--CDM model one finds that $Om(z)=\Omega_{m0}$  so only if the dark energy is given by a cosmological constant the $Om(z)$ remains constant. For  our model we find that $Om(z)$ can be written as
\be
\n{om2}
Om(z)=\frac{\Big\{B(1+z)^{3}+ (1-B)(1+z)^{3(\nu+1)}\Big\}^{\frac{1}{\nu+1}}-1}{(z+1)^{3}-1},
\ee
Now, we are going to analyze some important regimes. In the far  future, $z\rightarrow -1$,  Eq.(\ref{om2}) reaches  the value 1. For the present time, $z\rightarrow 0$, the Eq. (\ref{om2}) gives $[B+(1-B)(\nu+1)]/ (\nu+1)$ which is positive for $\nu>0$. In addition, using the constrained values of $(\nu,\Omega_{x0})=(1.19, 0.61)$ coming from  the $H_{obs}$ data, we plot the evolution of $Om(z)$ for both the MHRDE and the $\Lambda$--CDM models [see Fig.(\ref{fig8})]. Interestingly enough, in the interval $z \in [0, 1)$ the $Om_{MHRDE}>\Omega_{m0}$ the inequality is reversed at redshift, $z\geq 1$, as it happens with a phantom dark energy model. Thus the same behavior occurs with a quintessence model \cite{om}. 

\vspace{0.5cm}

As a closing comment, we would like to address a discussion concerning the values of $\al$ and $\beta$ taken into account through this section. Here we have focused on the transition of the Universe between a stage dominated by dark matter followed by an  era  dominated by the holographic dark energy that  makes the Universe exhibt an accelerated expansion (present-day scenario) and, in both stages  a nonlinear interaction in the dark sector has been taken into account. In order to estimate the parameter $\nu$ and $\Omega_{x0}$ we have used the values of $\al$ and $\beta$ which are consistent with the $\chi^{2}$-statistical analysis because  they fulfill the condition $\chi^{2}_{dof}<1$. Now,   we are going to explore a modification on the aforesaid model  by adding a third component, say $\ro_{m}$, which does not interact with $\ro_{c}$ and $\ro_{x}$. The total energy density reads as  $\ro_{t}=\ro_{m}+ \ro$ with $\ro=\ro_{c}+\ro_{x}$ and  the MCE  is split  as
\be
\n{nc1}
\ro'+ \al\ro-(\al-\beta)\ro_{x}=0,
\ee
\be
\n{nc2}
\ro_{m}'+ \al\ro_{m}=0
\ee
so (\ref{nc1}) is equal (\ref{06})  while (\ref{nc2}) shows that the third component is decoupled from the interacting dark sector, indicating that $\ro_{m}$ feels the presence of the dark sector through the gravitational expansion of the Universe. From (\ref{nc2}) one finds that $\ro_{m}=\ro_{m0}a^{-3\al}$ whereas the behavior of dark matter and dark energy with the scale factor  can be obtained from  (\ref{cI}) and (\ref{xI}), respectively.   Using (\ref{cI}), (\ref{xI}) and (\ref{nc2}) one gets the Hubble parameter in term of the redshift $x=1+z$ and the relevant cosmological parameters

\be
\n{Ht2}
\frac{H(z)}{H_0}=\Big[(1-\Omega_{m0})\big(Bx^{3}+ (1-B)x^{3(\nu+1)}\big)^{\frac{1}{(\nu+1)}}+\Omega_{m0}x^{3\alpha}\Big]^{1/2}
\ee
\be
\n{B2}
B[\theta]=\frac{\nu+1}{\nu}\Big[\frac{\al-\beta}{1+\frac{1-\Omega_{x0}-\Omega_{m0}}{\Omega_{x0}}} + (1-\al)\Big]
\ee
where the flatness condition $1=\Omega_{x0}+\Omega_{c0}+\Omega_{m0}$ has been used. In what follows we  would to examine two  traits  of the model. First, we fix $\al=4/3$ to get a radiation contribution in the total density because it will address the problem of the dark energy at early times; thus,  as is well known the fraction of dark energy in the radiation era should fulfill the stringent bound $\Omega_{x}(z\simeq 1100)<0.1$ in order for the model be consistent with the big bang nucleosynthesis (BBN) data. For the priors $(H_{0}=72.2, \nu=1.19, \Omega_{x0}=0.7, \al=4/3)$ the Hubble data give as the best--fit values $\beta=0.1$ and $\Omega_{m0}=1.7\times 10^{6}$ along with $\chi^{2}_{dof}=0.78< 1$ with a fraction of dark energy $\Omega_{x}(z\simeq 1100)=0.2$ nearly close to the BBN's bound. Second, employing the Hubble data for (\ref{Ht2}) we estimate the best--fit value of $\Omega_{m0}$ and $\alpha$. Taking as priors  $(H_{0}=72.2, \nu=1.19, \Omega_{x0}=0.61, \beta=0.15)$ the $\chi^{2}$--analysis yields as the best--fit values $\al=1.01$ and $\Omega_{m0}=9.9\times 10^{-5}$ together with a $\chi^{2}_{dof}=0.79< 1$. Moreover, the latter case leads to an early  dark energy  $\Omega_{x}(z\simeq 1100)=0.01<0.1$ which is consistent with the bounds reported in \cite{EDE1} or with the future constraints achievable by Planck and CMBPol experiments \cite{EDE2}.  Therefore, taking the third component as the radiation term or a nearly radiation contribution, has helped to validate the first model,  indicating that the value of the cosmological parameters selected are consistent with BBN constraints.

\section{conclusion}
We have investigated the interacting dark sector with a MHRDE, where the IR cutoff  is provided by the modified Ricci scalar in a form of a linear combination of $\dot{H}$ and $H^{2}$, so that $\ro_x=2(\dot H + 3\al H^2/2)/(\al -\bb)$. This appealing cutoff yields a MCE in the dark sector with constant barotropic indexes, identified with $\al$ and $\beta$. We have introduced an interaction between the dark matter and dark energy densities, homogeneous of degree 1 in the variables $\ro$ and $\ro'$, and solved the source equation for the total energy density of the mix. The equation of state of this effective fluid is that of the relaxed Chaplygin gas. The latter interpolates between a matter dominated phase at early times and an accelerated expanding phase dominated by the MHRDE at late times. 

We have used the observational Hubble data to constrain the cosmological parameters of the model and to compare with the $\Lambda$--CDM model.  Taking as a reference point  $(\al_{f},\beta_{f})=(1.01, 0.15)$  we get the best fit at $\theta_c=(\nu,\Omega_{x0})=(1.19, 0.61)$ with $\chi^2_{min}=7.86$ leading to a good fit with $\chi^2_{dof}=0.786 <1$ per degree of freedom [see Fig.(\ref{fig1})]. We have established that a model with a holographic dark energy $\ro_{x} \propto R$ leads to $0.59<\Omega_{x0}<0.69$ which is close to the bounds $\Omega_{x0}=0.73$ provided by  WMAP-7 \cite{WMAP7}. For $\beta=0.1$, we have shown that the Ricci cutoff ($\alpha=4/3$) is  consistent with other values of $\alpha$ because  it fulfills the goodness condition ($\chi^2_{dof}<1$). In addition, we have obtained the allowed range of $(\nu,\Omega_{x0})$ when one varies $\al$ and $\beta$  [see Table.(\ref{VP})]. Properly estimating the $H_{0}$ and $\Omega_{x0}$ with the Hubble data we have confronted the $\Lambda$--CDM  with  the MHRDE model; thus, both models give some bounds of the pair  $(H_{0}, \Omega_{x0})$ consistent the those reported in \cite{WMAP7}. Besides, we have taken into account the SNe Ia with the Union2 data for calculating the best--fit values of $\nu$ and $\Omega_{x0}$. It led to $\nu=1.5$ and $\Omega_{x0}=0.70$ with $\chi^2_{dof}=0.812 <1$  while the Hubble data gave $\nu=1.19$ and $\Omega_{x0}=0.61$ with  $\chi^2_{dof}=0.786$. 

We have found that the equations of state of the dark energy equation and the unified fluid, at the best--fit values $\theta_c$, do not cross the phantom divide line [see  Fig.(\ref{fig5})] while the present value of the equation of state for the dark energy is $\omega_{x0}=-0.88$.

From the deceleration parameter [see Fig.(\ref{fig6})] and the best fit values $\theta_c$, we have obtained that the acceleration starts at $z_{acc}=1.06$ hence, the model with a MHRDE enters the accelerated regime earlier than the $\Lambda$--CDM with $z_{acc}=0.75$. We have shown that the density parameters $\Omega_c$, $\Omega_x$, and its ratio $r(z)$ in  Fig.(\ref{fig7}) seem to alleviate the coincidence problem. It is related to the decreasing behavior of the interaction with the redshift and its vanishing limit in the far future [see  Fig.(\ref{fig7})]. We have applied the geometrical  $Om$  diagnostic method to our model which seems include several aspects different from the $\Lambda$--CDM model.

Finally, we have analyzed a new model with a third fluid decoupled from the dark interacting sector. The latter scheme allowed us to examine the issue of the early dark energy at $z \simeq 1100$ as well as  to check the match of the models with the BBN constraints \cite{EDE1}, \cite{EDE2}.


\acknowledgments
We would like to thank the referee for making useful suggestions, which  helped improve the article.
LPC thanks  the University of Buenos Aires  for their support under Project No. X044 and the Consejo Nacional de Investigaciones Cient\'{\i}ficas y T\' ecnicas (CONICET) through the research Project PIP 114-200801-00328. MGR is partially supported by CONICET.

\vspace{1cm}

\end{document}